\documentclass[preprint,onecolumn, amssymb, showpacs]{revtex4-1}				  
\usepackage{graphicx}			
\usepackage{dcolumn}			
\usepackage{bm}					
\usepackage{float}			
\usepackage{amsmath,amsfonts}	
\usepackage{url}					
\usepackage{setspace}		
\usepackage{lineno}   			
\usepackage{color}
\usepackage{graphicx} 
\usepackage{hyperref}
\usepackage{cleveref}
\usepackage{natbib}

\begin{document} 

\title{Graded metasurface for enhanced sensing and energy harvesting}

\author{Jacopo M. De Ponti}%
\email[Corresponding author ]{e-mail: jacopomaria.deponti@polimi.it}
\affiliation{Dept. of Civil and Environmental Engineering, Politecnico di Milano, Piazza Leonardo da Vinci, 32, 20133 Milano, Italy}
\affiliation{Dept. of Mechanical Engineering, Politecnico di Milano, Via Giuseppe La Masa, 1, 20156 Milano, Italy}
\author{Andrea Colombi}
\affiliation{Dept. of Civil, Environmental and Geomatic Engineering, Stefano-Franscini-Platz 5, 8093 Z\"urich, Switzerland}
\author{Francesco Braghin}
\affiliation{Dept. of Mechanical Engineering, Politecnico di Milano, Via Giuseppe La Masa, 1, 20156 Milano, Italy}
\author{Raffaele Ardito}
\author{Alberto Corigliano}
\affiliation {Dept. of Civil and Environmental Engineering, Politecnico di Milano, Piazza Leonardo da Vinci, 32, 20133 Milano, Italy}
\author{Richard V. Craster}
\affiliation{Dept. of Mechanical Engineering, Imperial College London, South Kensington Campus, London, U.K.}
\affiliation{Dept. of Mathematics, Imperial College London, South Kensington Campus, London, U.K.}
\date{\today}


\begin{abstract}
\noindent In elastic wave systems, combining the powerful concepts of resonance and spatial grading within structured surface arrays enable resonant metasurfaces to exhibit broadband wave focusing, mode conversion from surface (Rayleigh) waves to bulk (shear) waves, and spatial frequency selection. Devices built around these concepts allow for precise control of surface waves, often with structures that are subwavelength, and utilise rainbow trapping that separates the signal spatially by frequency. Rainbow trapping yields large amplifications of displacement at the resonator positions where each frequency component accumulates. We investigate whether this amplification, and the associated control, can be used to create energy harvesting devices; the potential advantages and disadvantages of using graded resonant devices as energy harvesters is considered. 

\noindent We concentrate upon elastic plate models for which the $A_0$ mode dominates, and take advantage of the large displacement amplitudes in graded resonant arrays of rods,  to  design innovative metasurfaces that focus waves for enhanced piezoelectric sensing and energy harvesting. Numerical simulation allows us to identify the advantages of such graded metasurface devices and quantify its efficiency, we also develop accurate models of the phenomena and extend our analysis to that of an elastic half-space and Rayleigh surface waves. 

%
%

\end{abstract}

\maketitle
\section {Introduction} 
\label{sec:intro}
\noindent Metamaterials, in their modern guise, emerged about two decades ago in optics \citep{pendry}  and their potential in creating artificial media, with properties not found in nature, and the consequent opportunities in wave physics, have triggered intense research activity. Metamaterial concepts have now become a paradigm for the control of waves across much of physics and engineering and are widely used in   
 electromagnetism \citep{pendry,wiltshire2004,sar12008,werner2016}, acoustics \citep{Liu,craster2012}  and elasticity \cite{elastic_book}. 

\noindent Whilst many theoretical studies have made the basic concepts well-established, manufacturing complex 3D bulk metamaterials remains a challenge. Attention has therefore focussed on metasurfaces because wave propagation is often dominated by scattering from, transmission through, or waves guided along, surfaces;  surfaces are also easier to pattern, print, or manufacture. 
Metamaterials are available in a variety of shapes and sizes, but all feature arrangements of  local heterogeneities are often repeated periodically.

\noindent In elasticity, metamaterials achieved popularity with ideas based around Bragg scattering, phononic crystals, and material contrast to create band-gaps \cite{brule,mouldin_waves,djafar} that often draws upon ideas from the photonic crystal community. In geophysical settings, and for applications in Mechanical and Civil engineering, there has been a drive to obtain broader band-gaps at a low frequency \cite{achaoui_seismic,miniaci_seismic,DAlessandro,DAlessandro1} and to use locally resonant metamaterials \citep{eleni2,finocchio,carta_giorgio}.  Locally resonant metamaterials are characterised by resonating elements, analogous to Helmholtz resonators \cite{cocacola} in acoustics and Fano resonances \cite{fano}, enabling the creation of low-frequency band-gaps in structures relatively small compared to the wavelength;  
these ideas can be augmented by tuning or spatially grading the resonators to broaden or decrease band-gaps. 
\noindent At a smaller scale, and consequently higher frequency, wave redirection and protection \citep{ColombiSensing,andrea6,Matlack26072016,celli1} promise  advantages in reducing vibrations in manufacturing, or in laboratories, where high tolerances or precision measurements (e.g. interferometry) are required; similar motivations occur in ultrasonic inspection to increase the signal to noise ratio. To capitalize on these recent metamaterial designs, that can improve energy focusing over a wide frequency band,   energy harvesting is another  attractive application; at the millimetric, or centimetric, lengthscales concerned these would operate at frequencies in the kHz range.

\noindent Fuelled by demands to reduce the power consumption of small electronic components, vibration-based energy harvesting has received considerable attention over the last two decades; it is attractive to power devices using existing vibrational energy that reduces, or removes, costs associated with periodic battery replacement and the chemical waste of conventional batteries. Many applications arise: wireless sensor networks for civil infrastructure monitoring, unmanned aerial vehicles, battery-free medical sensors implants, and long-term animal tracking sensors \cite{ErturkHarvesting}. 
Among the various possible energy harvesting methods, piezoelectric materials offer several advantages due to their large power densities and ease of application \cite{ErturkPiezoBook,StevenAnton,RiazAhmed}. Piezoelectric energy harvesting is driven by the
deformation of the host structure due to mechanical or acoustic vibrations that convert to an electrical potential via embedded piezoelectric materials. 
\noindent To increase harvester efficiency, using ideas based around structuring surfaces, several approaches such as creating a parabolic acoustic mirror, point defects in periodic phononic crystals and acoustic funnels have been employed \cite{Carrara,HangHarvesting,LiangWu,Gonnella}; lenses to concentrate narrow band vibrations have been proposed using phononic crystals \cite{Tol, Tol2, 
AhmadLens} and resonant metamaterials endowed with piezoelectric inserts have appeared very recently \cite{Sugino}. Our aim here is to complement these studies by using a graded array to create a metawedge \cite{colombi16a}, and introduce piezoelectric elements into the array, this addresses one of the main challenges in energy harvesting which is to achieve broadband energy production from ambient vibration spectra. At present, multimodal harvesting, or exploiting non-linear behaviour, partly addresses this challenge \cite{ErturkHarvesting}. Multimodal harvesting requires multiple bending modes with close, and effective, resonant peaks thereby leading to a broadband device; these systems can be affected by two problems: low overall power density (power/volume or power/weight) and complex interface circuits \cite{ErturkHarvesting}. Similarly to harvesting, resonant metamaterial devices have also been affected by a limited bandwidth. However recent research efforts have proven how this can be enlarged by adopting nonlinear resonators \cite{CVETICANIN201789,carbonara} and, above all, graded array designs \cite{colombi16a}. 

\noindent In this article, our approach is to use graded resonator arrays to concentrate energy exploiting the properties of the rainbow trapping device shown in Fig.~\ref{Harvester}. Because such systems already contain a collection of resonators, the inclusion of vibrational energy harvesters is straightforward leading to truly multifunctional meta-structures combining vibration insulation with harvesting. 
\noindent In section \ref{sec:metawedge} we recall the resonant metawedge design principles for an elastic plate and elastic halfspaces. Section \ref{sec:harvester_design} presents the harvester design and we show how it can be tuned by altering the rod length, grading angle and spacing in order to maximise the focussing. Here we also introduce a simplified analytical model for the harvester so that optimal design parameters are rapidly found. 3D numerical simulations of the electromechanical device installed on a plate and on a half-space are presented in section \ref{sec:results}. Finally, we draw together some concluding remarks in section \ref{sec:conclude}. 
\section{The graded resonant metasurface for elastic waves}
\label{sec:metawedge}

\noindent A cluster of rods attached to an elastic substrate whether acting as a phononic crystal, for short pillars \citep{wu,pennecchia,younes2011}, or as a resonant metamaterial, for long rods \cite{matthieu}, creates a versatile system for controlling elastic waves and lends itself well to the fabrication of graded designs. The physics of the resonant metasurface is described through a Fano-like resonance \citep{fano}; a single rod attached to an elastic surface readily couples with the motion of either the $A_0$ mode in a plate, or the Rayleigh wave on a thick elastic substrate (half-space),  and the coupling is particularly strong at the longitudinal resonance frequencies of the rod. Mathematically, the eigenvalues of the equations describing the motion of the substrate and the rod are perturbed by the resonance, mode repulsion occurs, and complex roots arise leading to the formation of a band-gap \citep{perkins86a,landau58a}. When the resonators are arranged on a subwavelength cluster (i.e. with $\lambda$, the wavelength, $\gg$ than the resonator spacing), as in the metasurface we use, the resonance of each rod acts constructively enlarging the band-gap until, approximately, the rod's anti-resonance \citep{matthieu,andrea_tree}. Because the resonance frequency of the rod, which depends on rod height, determines the band-gap position, then by simply varying the length of the rods one gets an effective band-gap that is both broad and subwavelength; the rod length is therefore a key parameter for metasurface tunability. However both periodicity and/or the distribution of the rods cannot be overlooked as they also shape the dispersion curves leading to a zone characterised by dynamic anisotropy and negative refraction \cite{kaina15a,ColombiMechRew}.  
\noindent The band-gap frequency $f$ is calculated approximately starting from the resonator length, $h$, and the well-known formula \cite{ColombiMechRew,ColombiSensing}:
\begin{eqnarray}
\label{eqn:AxialResonance}
f=\frac{1}{4 h}\sqrt{\frac{E}{\rho}},
\end{eqnarray} 
where  $E$ is the Young's modulus of the material and $\rho$ its density.  Eq. (\ref{eqn:AxialResonance}) holds provided the substrate is sufficiently firm so that one end of the rod is effectively clamped. Conversely, an overly soft substrate (e.g. a very thin plate) may result in a rod behaving as a rigid body connected with a spring thus compromising the amplification. In practical terms, this provides bounds on the plate thickness and requires cross-checking, via finite element simulations, the modal shape of the longitudinal eigenmode. Similar considerations are valid for the half-space, that, again, cannot be too soft when compared to the resonators \cite{andrea_tree}.

\begin{figure}[H]
\centering
    \includegraphics[width=1.05\textwidth]{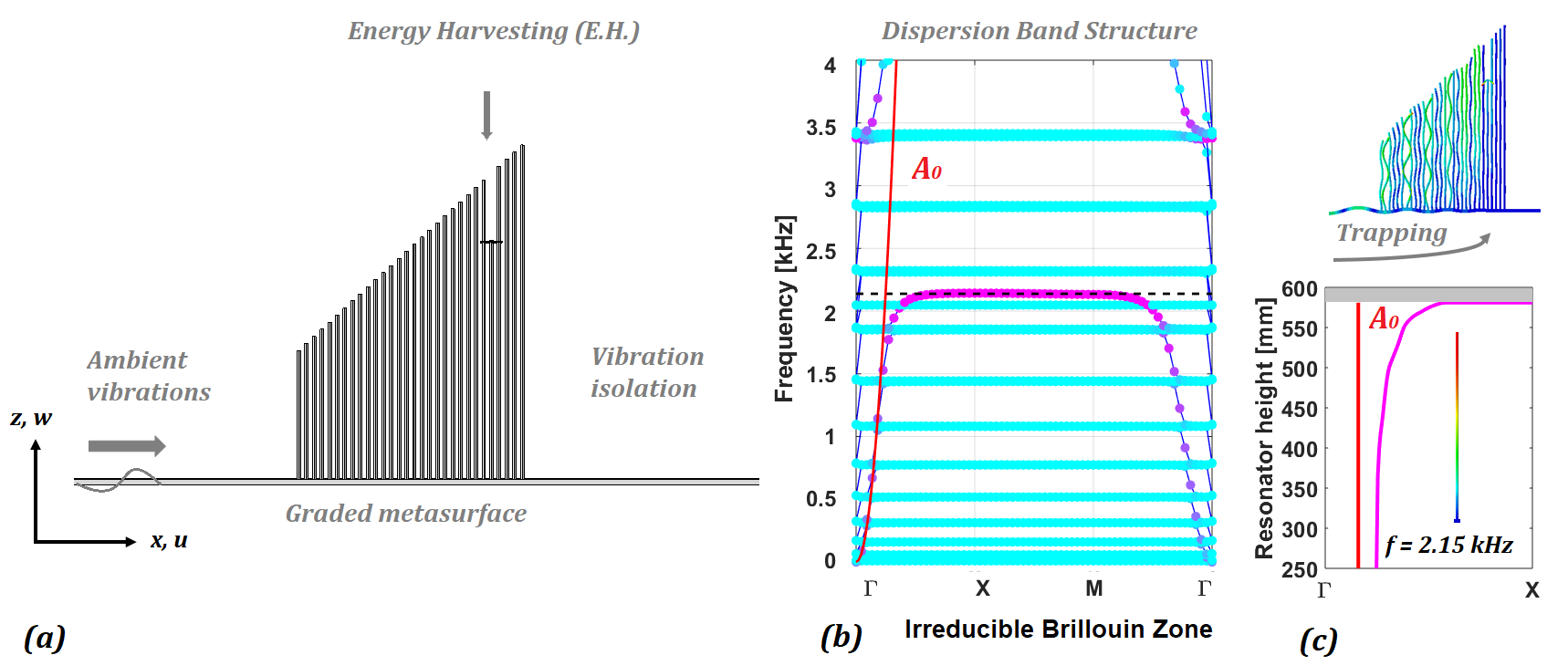}
    \caption{The elastic metasurface used for energy harvesting (a) and conventional dispersion curves (b) for a periodic array of identical resonators of fixed height (here we take the resonator to have height 581~mm, just before that endowed with piezo-patches in the graded wedge of (a)). Panel (c) illustrates the 
 trapping mechanism for the axial mode due to the metasurface grading by showing the relevant dispersion curve at fixed frequency but varying resonator height; the rod is shown inset at 581~mm.}
\label{Harvester}
\end{figure}

\noindent A schematic showing the resonant metawedge is depicted in Fig. \ref{Harvester}(a). We place a graded array of rods atop an elastic plate, and these rods have resonances (both flexural and longitudinal); if we assume the rods have constant height then these naturally appear in the dispersion curves of Fig.~\ref{Harvester}(b). These dispersion curves are  computed along the 2D irreducible Brillouin zone \cite{kittel} using finite elements, \cite{comsol} in a physical cell containing a single resonator, that incorporate the Bloch phase shift  
via  Bloch-Floquet periodic boundary conditions.   Fig.~\ref{Harvester}(b) shows the 
 longitudinal resonance (dashed line), and additionally a large number of flexural resonances are also visible. The impact of these flexural resonances on the wave  propagation is negligible \cite{williams}, at least in terms of the band-gaps. 
An unconventional dispersion curve representation, where the frequency is fixed and the resonator height varied, as in Fig. 
 \ref{Harvester}(c), shows how rod height affects the array. By grading the array from short to tall resonators, Fig. 
 \ref{Harvester}(c),  we see that the dispersion curve hits the bottom of the band-gap and subsequently the energy must be reversed and propagate back through the array;   equally importantly the group velocity tends to zero at this turning point and the wave slows down and therefore spends time in the vicinity of the resonator feeding energy into it. 

\noindent The resonant metawedge can also be attached onto a deep elastic substrate where it is capable of mode conversion or the trapping and reflection of Rayleigh waves. The effect is underpinned by the hybridisation between the longitudinal resonance of the rods, and the vertical component of Rayleigh waves \cite{andrea_tree,matthieu,colombi16a}; the type of interaction, conversion or trapping, depends on the direction of wave propagation across the metawedge. When moving towards increasing rod lengths we obtain trapping, conversely by reversing the wave direction, we obtain Rayleigh to shear, S, wave conversion. Because each rod has a different resonant frequency, the turning point (where conversion or trapping take place) varies spatially according to the frequency content of the incident wave (the so-called rainbow effect \cite{hess2007}).  Trapping is by far the most interesting property for energy harvesting as it leads to higher amplification on the resonator responsible for the trapping.  For the elastic plate, as the substrate, there is no conversion to a bulk mode as this requires a deep substrate,  but the concept of trapping carries across to the antisymmetric $A_0$  Lamb mode.

\section{Harvester design}
\label{sec:harvester_design}
\noindent Concentrating energy at a known spatial position, as the graded array does, is only part of the requirement for harvesting: It is also necessary to design an arrangement for the piezoelectric patches that takes full advantage of the displacements induced within the array and upon the rods. To understand the mechanisms involved we will take a single piezo-augmented resonator and place it within the graded array (Fig.~\ref{Harvester}).

\begin{figure}[H]
\centering
    \includegraphics[width=1\textwidth]{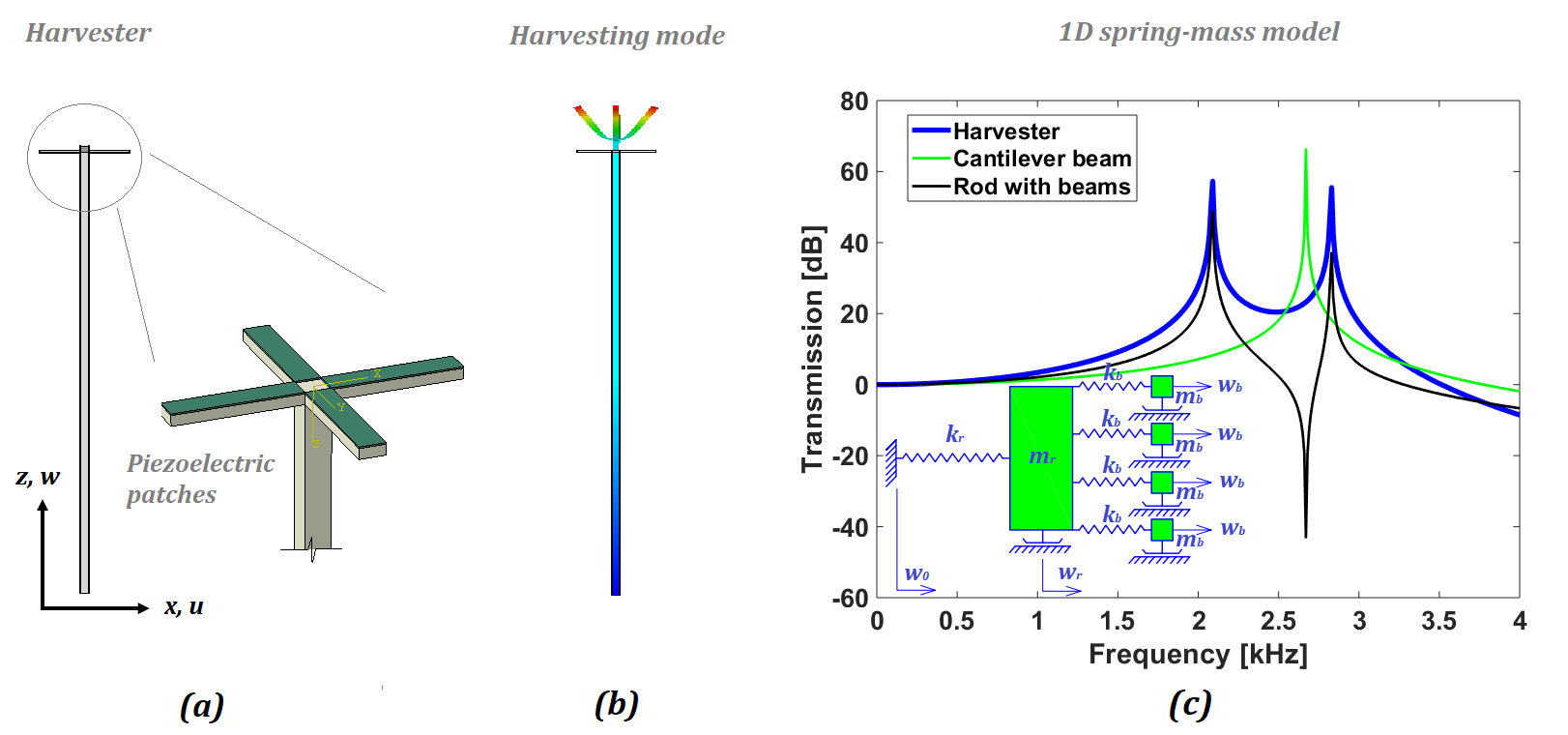}
    \caption{Harvester shown with the piezoelectric patches (a), the harvesting fundamental mode (b) showing the double amplification occurring when the cantilever beams fundamental mode and the rod axial mode are tuned to operate together. (c) shows the transmission spectrum for a single beam, the simple spring-mass model (shown inset) of a rod with beams and analytical transmission of the harvester.}\label{Harvester_patches}
\end{figure}

\noindent Recalling that the dominant mode of interest is the longitudinal one, the harvester we use is the rod, with four cantilever beams arranged in a cross-like shape placed upon the top, as depicted in Fig.~\ref{Harvester_patches}(a); each beam embeds a piezoelectric patch and their motion in harvesting mode is shown in Fig.~\ref{Harvester_patches}(b). In our elastic modelling we consider the plate and rods to be made from 
 aluminium ($E_a=70$ GPa, $\nu_a=0.33$ and $\rho_a=2710$ kg$/$m$^3$) while the piezoelectric patches are made of PZT-5H ($E_p=61$ GPa, $\nu_p=0.31$ and $\rho_p=7800$ kg$/$m$^3$).

\noindent The harvester design,  Fig.~\ref{Harvester_patches} (a), has strong dynamic coupling between the rod and the cantilever beams as we carefully design it to work in the double amplification regime coupling the axial fundamental mode with the flexural one of the beam (see Fig.~\ref{Harvester_patches} (b)) i.e. the lengths of the cantilever beams are not arbitrary. 
 Fig. \ref{Harvester_patches} (c) illustrates the transmission spectra for a single cantilever beam, the harvester we use and a simple although very accurate model of a rod with beams and enables us to rapidly tune the system. 
 
\noindent Provided the axial and flexural frequencies are matched, an amplification over a wide frequency range is obtained. We design the harvester to maximize this interaction at the trapping frequency of the metasurface in a range between $2$ and $2.4$ kHz. 
We use an optimization procedure that maximizes the integral of the transmission spectrum in this frequency range subject to an area constraint given by the geometry of the piezoelectric patch (defined on the basis of commercial values) whose width and length are the same as the beam. The values are: $A_r=25$ mm$^2$ (square cross section), $A_b=10$ mm$^2$ (same width as the rod, i.e. 5mm), $l_r = 460$ mm, $l_b=25$ mm and $h_p = 0.3$ mm (the subscript r denotes the properties of the rod, b of the beams and p of the patches). This harvester design provides a multimodal response of the system, always exploiting the fundamental flexural mode of the beams.

\begin{figure}[H]
\centering
    \includegraphics[width=0.95\textwidth]{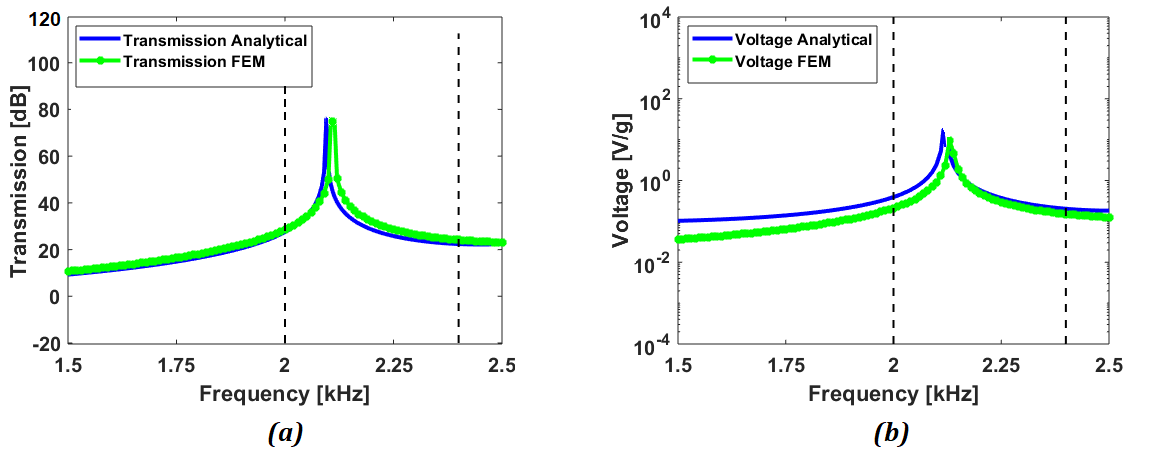}
    \caption{Comparison between the analytical and numerical transmission spectrum in short circuit (a) and voltage in open circuit (b).}
\label{Comparison}
\end{figure}

\begin{figure}[H]
\centering
    \includegraphics[width=0.95\textwidth]{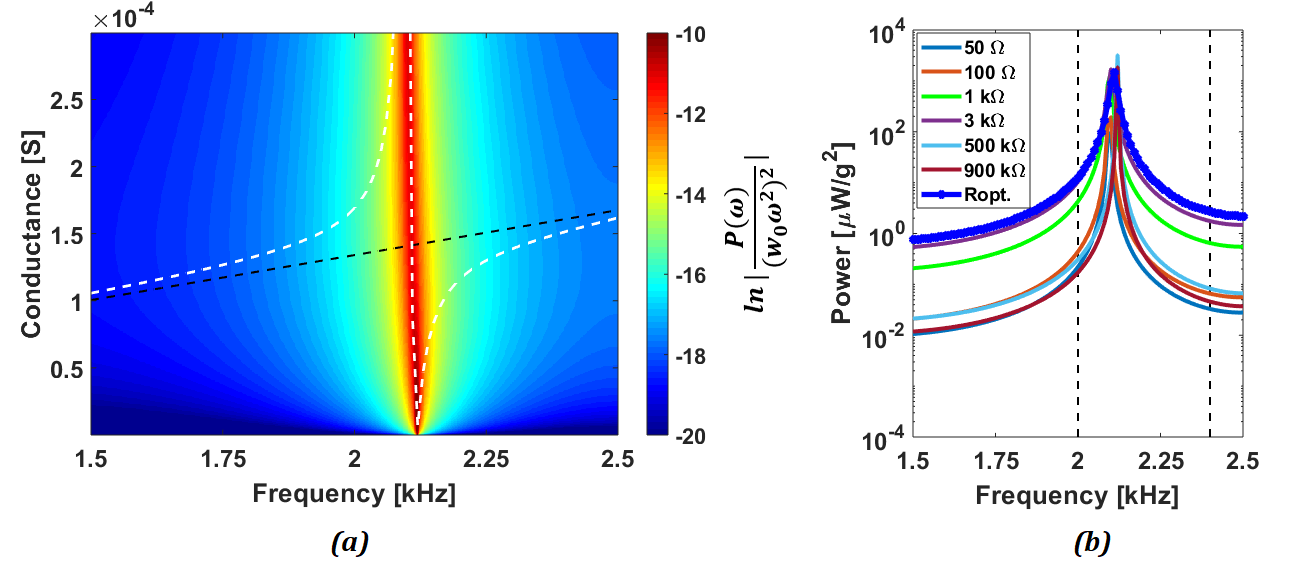}
    \caption{Power output vs. electrical conductance and frequency (a). Dashed white line shows the optimal loading at each excitation frequency. Dashed black line the cutoff frequency for different values of conductance. (b) show the corresponding electrical resistance for optimal power (normalized with respect to the gravitational acceleration $g^2$) (blue curve) compared with other resistance values.}\label{OptimalPower}
\end{figure}

\noindent The fundamental mode of the cantilever beam is that most suited for energy harvesting as it does not result in charge cancellation and has the highest power production;  with higher mode numbers this would decay by around two orders of magnitude \cite{ErturkPiezoBook, ErturkHarvesting}. In addition, this configuration provides high amplification at low frequency (relevant for ambient vibrations) due to both the slenderness of the rod and the added mass from the four cantilever beams. 
\noindent Tuning the cantilever beams to couple optimally with the fundamental mode of the rod leads to a rather lengthy numerical exercise, this is side-stepped here by using a simple, yet highly effective, mass-spring model. We outline the model below and note its accurate predictions in Fig. \ref{Comparison} in the frequency range of interest. 
We use the 1D spring-mass model, inset in Fig.\ref{Harvester_patches}(c), and adopt Timoshenko beam theory. Defining with $w_r(z,t)$ and $w_b(x,t)$ the vertical displacement of the rod and the beam respectively,
the effective mass is obtained writing the kinetic energy of the system for both the axial and flexural mode:
\begin{flalign}
&T_r(t) = \frac{1}{2}M_b\left[\frac{\partial {w_r(z,t)}}{\partial{t}}|_{z = l_r}\right]^2+\frac{1}{2}\int_{z=0}^{z=l_r}\rho_a A_r\left [\frac{\partial {w_r(z,t)}}{\partial{t}}\right]^2 dz = \frac{1}{2}\left[M_b + \frac{1}{3}\rho_a A_r l_r\right]\dot{w}_r^2(t)&\\
&T_b(t)= \frac{1}{2}\int_{x=0}^{x =l_b} (\rho_a A_b + \rho_p A_p) \left[\frac{\partial {w_b(x,t)}}{\partial{t}}\right]^2 dx = \frac{1}{2}\left[\frac{33}{140} (\rho_a A_b + \rho_p A_p) l_b\right] \dot{w}_b^2(t)&&
\label{EffectiveMass}
\end{flalign}
while elastic and electric lumped coefficients directly follow from the internal energy definition:  
\begin{flalign}
&U_r(t)=\frac{1}{2}\int_{z=0}^{z=lr}\int_{A}\left(T_{zz}S_{zz}\right) dA dz=\frac{1}{2}k_r w_r(t)^2&\\
&U_b(t) = \frac{1}{2}\int_{x=0}^{x=l_b}\int_{A}\left(T_{xx} S_{xx} + T_{xz} S_{xz} - D_{zz}E_{zz}\right) dA dx=\frac{1}{2}k_b w_b^2(t)-\theta w_b(t)v_b(t)-\frac{1}{2}C_p v_b^2(t)&&
\label{InternalEnergy}
\end{flalign}
with a separation of variables taken as: $w_r(z,t)=w_r(t){z}/{l_r}$ and $w_b(x,t) = \frac{1}{2}[3({x}/{l_b})^2-({x}/{l_b})^3]w_b(t)$, 
$M_b= 4(\rho_a A_b+\rho_p A_p)$  the total mass of the four beams, $\underline{T}$ and $\underline{S}$ the stress and strain fields, $\theta$ the electromechanical coupling coefficient and $C_p$ the internal piezoelectric capacitance. 
Lumped coefficients are then defined as: $k_r = {E_a A_r}/{l_r}$, $k_b = {1}/({\frac{l_b^3}{3E_a I_b}+\frac{l_b}{G_a A^*_b}})$, $\theta = -\frac{3e_{31}a_p}{2h_pl_p}(h_p^2+2b_bh_p-2y_nh_p)$, $C_p=\bar{\epsilon}_{33}a_pl_p/h_p$, $
m_r = {1}/{3} \rho_a A_r l_r + M_b$, $
 m_b = ({33}/{140}) (\rho_a A_b +\rho_p A_p)l_b$ and 
 $G_a = E_a/2(1+\nu_a)$ being the aluminium shear modulus, $a_p$ the patch width, $b_b$ the beam thickness, $y_n$ the neutral axis of the composite beam, $e_{31}$ the considered piezoelectric coefficient and $\bar{\epsilon}_{33}$ the constant-stress dielectric constant.   
The dynamics, of the electromechanical problem, are now succinctly described by three linear coupled ordinary differential equations:
\begin{eqnarray}
\begin{cases}
m_r\ddot{w}_r+k_r(w_r-w_0)+4k_b(w_r-w_b)=0
\\ 
 m_b\ddot{w}_b+k_b(w_b-w_r)-\theta v_b =0
\\ 
 C_p\dot{v}_b+\tilde{G}v_b+\theta \dot{w}_b = 0
\end{cases}
\label{CoupledEquations}
\end{eqnarray}
where $\tilde{G}$ is the electrical conductance.
 Fourier transforming (\ref{CoupledEquations}) gives the transfer function: 
\begin{eqnarray}
\tilde{T}(\omega) = \frac{k_r k_b}{(-m_r \omega^2 + k_r + 4k_b)(-m_b \omega^2 + k_b+\frac{i\omega \theta^2}{i \omega C_p + \tilde{G}})-4k_b^2}
\label{TransferFunction}
\end{eqnarray}
from which the voltage and power follow directly as:
\[
\tilde{V}(\omega)= -\frac{i\omega \theta}{i \omega C_{p} + \tilde{G}} \tilde{T}(\omega)w_0 , \qquad 
\tilde{P}(\omega)= \tilde{G}\left[\frac{i\omega \theta}{i \omega C_{p} + \tilde{G}} \tilde{T}(\omega)w_0\right]^2 
\]
with $w_0$ the imposed harmonic displacement amplitude. For brevity we do not incorporate damping, but it can be easily included in  (\ref{CoupledEquations}) by introducing a term linearly dependent on the velocity $\dot{w}$. The main contributor to damping in this system is provided by the piezoelectric material since the quality factor of aluminium is very high. However we neglected damping here since the thickness of the patch is very small with respect to that of the beam.
This simplified model provides  surprisingly accurate predictions in the frequency range of interest (see Fig. \ref{Comparison} (a) and (b)), with just a $0.7 \%$ of error in the natural frequency prediction.
\noindent The ratio $C_p/\tilde{G}$ (i.e. $RC_p$) defines the time constant of the circuit $\tau_{RC}$ providing measure of the time required to charge the capacitor through the resistor. The time constant is related to the cutoff circular frequency which is an alternative parameter of the $RC$ circuit:
\begin{eqnarray}
\omega_{RC}= \frac{1}{\tau_{RC}}=\frac{\tilde{G}}{C_p}
\label{cutoffFrequency}
\end{eqnarray}
The values of $\tilde{G}$ maximising the electric power at each excitation frequency are obtained by  imposing its stationarity 
 with respect to $\tilde{G}$ (the dashed white line in Fig. \ref{OptimalPower}(a)):
\begin{eqnarray}
\resizebox{.94\hsize}{!}{
$\tilde{G}_{opt.}(\omega) = \frac{k_r\omega \theta^2 + 4k_b\omega \theta^2 - m_r \omega^3 \theta^2 - C_pk_rm_b\omega^3 - C_pk_bm_r\omega^3 - 4C_pk_bm_b\omega^3 + C_pm_rm_b\omega^5 + C_pk_rk_b\omega}{-k_rk_b + k_rm_b\omega^2 + k_bm_r\omega^2 + 4k_bm_b\omega^2 - m_rm_b\omega^4}$
}
\label{OptimalConductance}
\end{eqnarray}
The optimal electric conductance is then obtained from the intersection of (\ref{cutoffFrequency}) and (\ref{OptimalConductance}) (the intersection of black and white dashed lines in Fig. \ref{OptimalPower} (a)) . The corresponding optimal electrical resistance is $R_{opt}=7.075 k\Omega$.

\noindent Now that we have obtained an harvester optimised for longitudinal rod resonance, it can be 
 introduced into the metasurface array to assess whether its performance is increased when rainbow trapping occurs. The considerable prior research with this metamaterial in the frequency range of interest \citep
{matthieu,andrea6} provide essential insights to the numerical simulations. 
    
\section{Numerical results}
\label{sec:results}
\noindent We present numerical simulations of the graded harvester on an elastic plate, the plate has the form of a finite width strip 
%
 to reduce computational complexity \cite{williams,colombi16a} whilst still preserving the physics of the resonant metasurface. The array is attached to an aluminium strip 30 mm wide and of thickness of 10 mm which is sufficiently stiff to avoid anomalous resonances in the rod. The array is composed of 30 rods, each with square cross section of area 25 mm$^2$, a linear height gradient ranging from 250 to 650 mm and a constant spacing between the rods of 15 mm; this results in a $\sim43^{\circ}$ slope angle.  Equation (\ref{eqn:AxialResonance})  suggests that the metasurface will have the longitudinal fundamental mode in the range 2 to 5 kHz and rod number 26 (with the rod numbering in the array having 1 the shortest resonator and 30 the longest), with length 460 mm, is the harvester designed in section \ref{sec:harvester_design}. 

\begin{figure}[H]
\centering
    \includegraphics[width=1\textwidth]{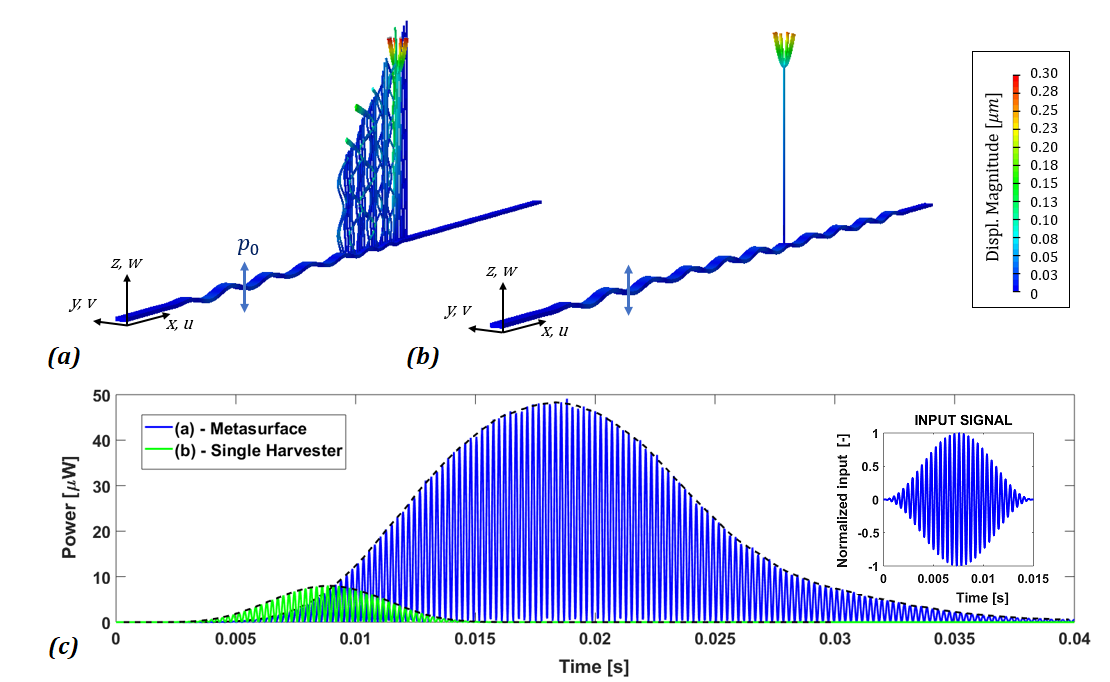}
    \caption{Harvester on a plate strip with (a), and without (b), the metasurface at time t = 10 $ms$. Electrical power for a maximum input acceleration of 1$g$ with, and without, the metasurface (c).
    \label{PlateFocusing1} }
\end{figure}

\noindent To benchmark the graded harvester, and to highlight the advantage of the elastic wave focussing, a time domain simulation with a Gaussian excitation centered at 2.15 kHz (frequency of the main mode of the harvester) is performed. This is compared to an isolated harvester on the same plate with, and without, the metasurface (see Fig. \ref{PlateFocusing1} (a) and (b)). 
\noindent Both models are excited with a vertical pressure $p_0$, on a surface $S_0=$ 650 $mm^2$, applied in the same position which generates an $A_0$ Lamb wave. The amplitude of the pressure is such that a maximum acceleration of 1$g$ is imposed at the input. FEM simulations on these 3D structures are performed in ABAQUS CAE 2018  introducing piezoelectric elements on the top of the harvester, with zero voltage boundary conditions on the bottom face of the piezoelectric patches. Electrodes are modelled introducing a voltage  constraint  on the piezo-faces in order to define equipotential surfaces. The electrodes on top are connected in series and attached to the previously defined optimal resistance; to model this, a customised Fortran subroutine has been implemented in ABAQUS, according to \citep{Gafforelli,
Gafforelli1}.  Absorbing boundary conditions are imposed with a parametric analysis assigning different values of damping at the finite elements along $x$ at the edge of the strip. Damping is computed adopting a cubic law function with $c_{max}$ $=$ $10^6$, according to \cite{AbsorbingBC}. 
It can be clearly seen that the metasurface (see Fig. \ref{PlateFocusing1} (c)) provides a strong amplification of the electrical power, and when the harvester is embedded in the metasurface,  due to  the lower group velocity of the waves interacting with the array grading, the power generation peak occurs at a later time. The wave velocity decreases along the grading, reaching the harvester with a nearly zero value. For short propagation time (e.g. 10 ms), the grading has a  detrimental effect on the power production. Conversely if the excitation is sufficiently long, energy stops in the position of the harvester loading it continuously and, in this case, the energy production peaks at about 20 ms; after this time, waves are backscattered and the power production decreases. It is worth noticing that not only is the peak higher, but globally the area below the curve is higher: this further demonstrates that energy is trapped inside the metasurface. 
We recall that the metasurface also provides strong vibration attenuation over the entire frequency range due to band-gap generated by the longitudinal resonances of the rods. The behaviour is strongly subwavelength, as can be noticed comparing the rod spacing with the wavelength of the $A_0$  mode in the frequency range we consider.\\
 
\begin{figure}[H]
\centering
    \includegraphics[width=1.04\textwidth]{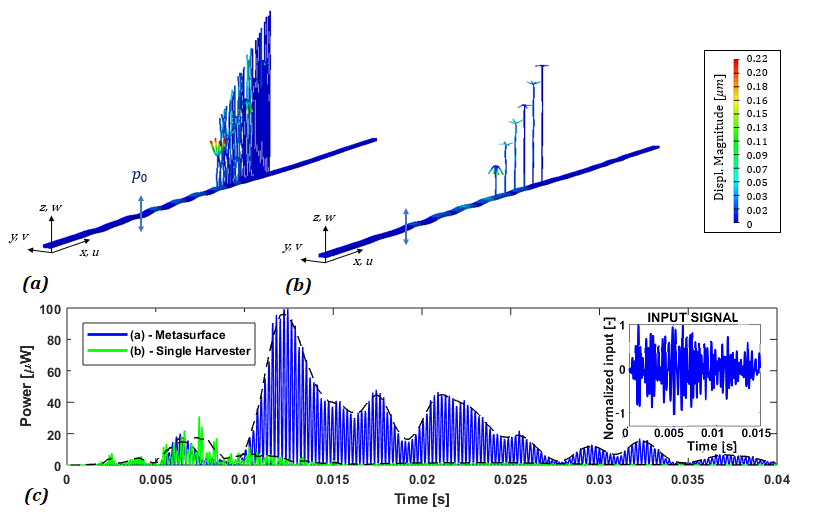}
    \caption{Six harvesters on a plate strip with (a) and without (b) the metasurface at time t = 10 $ms$. Electrical power for a maximum input acceleration of 1$g$ with and without the metasurface (c).}\label{PlateFocusing2}
\end{figure}

\noindent It is possible to increase the harvesting bandwidth  by introducing other harvesters in different positions. In this way, for broadband input, energy is almost uniformly trapped inside the grading and feeds different harvesters. The harvesters introduced maintain  the same beams on top and only differ in terms of the rod length. From a practical, and economical perspective, this solution is more  feasible since the piezo patches are equal and so is the optimal resistance $R_{opt}$. Theoretically, a beam grading induces  higher power production for the isolated harvesters, as explained in section \ref{sec:harvester_design}. However, since  the harvesters are electrically connected in series, the probability of out of phase responses and hence charge cancellation increases. Therefore, unless one aims at powering multiple devices accommodating different $R_{opt}$, keeping the same piezoelectric beam size leads to a simpler and more controllable system.
The rod length of the harvesters is defined adopting a linear variation starting from resonator 26, with the same metasurface angle. As for the previous case, the piezoelectric patches of each harvester are connected in series to the same electrical resistance. The input vibration is a zero mean coloured noise with a frequency content mainly in the metasurface range between 2 and 5 kHz. This is applied with a uniform pressure on the same surface $S_0$, such that a maximum 1$g$ acceleration is imposed at the input. The optimal electrical resistance in this case is obtained with a parametric analysis of the entire system. This means that (\ref{OptimalConductance}), derived for an harmonic regime, cannot be used here because of the dynamic interaction between harvester of different length, the broadband and transient nature of the input signal. By performing different numerical analyses, a value of 1.18 $k\Omega$ is adopted. The case with isolated harvesters is then compared with the one where they are introduced inside the metasurface (Fig.~\ref{PlateFocusing2} (a) and (b)).
As in the previous case, it can be clearly seen that the metasurface (Fig.~\ref{PlateFocusing2} (c)) strongly amplifies the electrical power. The global behaviour shown in Fig.~\ref{PlateFocusing2} (c) is really similar to the one reported in Fig. \ref{PlateFocusing1} (c). The longer time necessary to reach the power production peak points to the same energy trapping mechanism discussed for the single harvester. Finally, we remark a phase synchronisation across the harvesters enhanced by the presence of the grading that partially avoids charge cancellation when they are connected in series.

\begin{figure}[H]
\centering
    \includegraphics[width=1.04\textwidth]{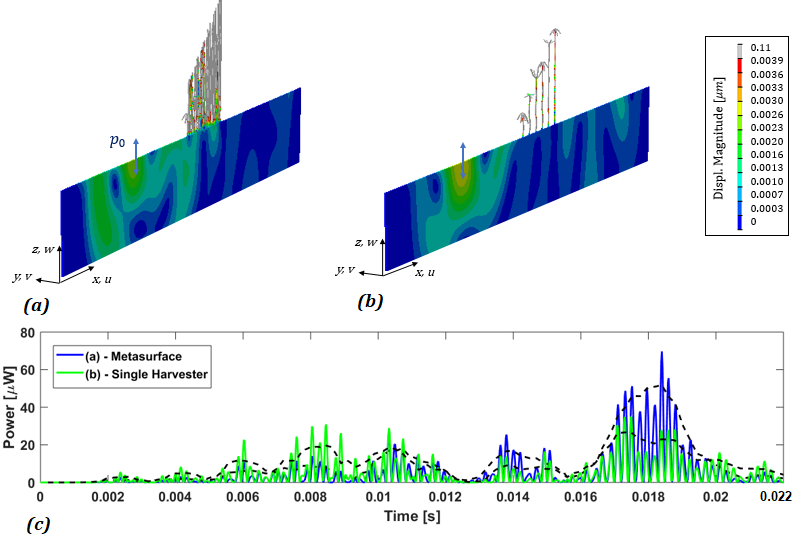}
    \caption{Six harvesters on an half-space strip with (a) and without (b) the metasurface (equal to the one on the plate strip) at time t = 10 $ms$. Electrical power for a maximum input acceleration of 1$g$ with and without the metasurface (c).}\label{Halfspace}
\end{figure}

\noindent The same mechanism is found when placing the same metasurface on an elastic half-space. As for the plate we took a finite width strip, but now of infinite depth with plane strain boundary conditions applied on these faces along the $y$ direction. This set-up represents the case where the metasurface is built directly on the ground or on a thick mechanical component to harvest ambient noise. We use the same coloured noise excitation adopted for the plate strip,  adopting a uniform pressure able to impose a maximum 1$g$ acceleration at the input. The same harvesters are placed inside, and outside, the metasurface (see Fig. \ref{Halfspace} (a) and (b)) and connected in series with an electrical resistance equal to the previous one (it is assumed to power the same device). The metasurface is able to provide a strong focusing due to rainbow trapping, inducing, at the same time, a Rayleigh wave band-gap (Fig. \ref{Halfspace} (a) and (b)). 
\section{concluding remarks}
\label{sec:conclude}

\noindent We have numerically demonstrated the advantage of combining graded resonant metamaterials with multiple piezoelectric harvesters both in terms of band widening and overall efficiency. This is demonstrated for both a plate, and a half-space, with a graded array  resonators; this exploits the wave trapping mechanism, increasing the performance of the introduced piezoelectric energy harvesters.  To exploit this behaviour we find that a sufficiently long excitation is required as the wave group velocity decreases along the grading reaching the value of zero in the position of the harvester. In this manner, before being backscattered, the waves enjoy a longer interaction with the harvester enhancing its power production when compared against the case of isolated harvesters. Additional features are that the system can be frequency-tuned simply by adding masses on the top of the rods and in this way it is possible to match the longitudinal mode of the rod and the flexural one of the piezoelectric patch even at very low frequencies; this is important from an application point of view where one wish to exploit actual low frequency ambient spectra ($\ll 1$ kHz).
Future work will act to address both the experimental verification of such a device and the integration  of non-linear elements in the  array or on the plate capable to exploit even lower frequency vibrations. 

\section*{Acknowledgements}
\noindent RVC thanks the UK EPSRC for their support through Programme Grant EP/L024926/1 and the Physics of Life grant EP/T002654/1. RVC also acknowledges the support of the Leverhulme Trust. Al.Cor. thanks the national project PRIN15 2015LYYXA8. An.Col. thanks the SNSF for their support through the Ambizione fellowship PZ00P2-174009. JMDP thanks Imperial College for its hospitality and Politecnico di Milano for the scholarship on "Smart Materials and Metamaterials for industry 4.0".


\newpage

 
\begin{thebibliography}{99}





\bibitem{pendry}
J.~B. Pendry, A.~J. Holden, D.~J. Robbins, and W.~J. Stewart.
\newblock Magnetism from conductors and enhanced nonlinear phenomena.
\newblock {\em IEEE Transactions on Microwave Theory and Techniques},
  47(11):2075--2084, 1999. 
   
\bibitem{wiltshire2004}
D.~R. Smith, J.~B. Pendry, and M.~C.~K. Wiltshire.
\newblock Metamaterials and negative refractive index.
\newblock {\em Science}, 305:788--792, 2004.

\bibitem{sar12008}
S.~A. Ramakrishna and T.~M. Grzegorczyk.
\newblock {\em Physics and applications of negative refractive index
  materials}.
\newblock CRC Press, 2008.

\bibitem{werner2016}
D.~H. Werner.
\newblock {\em Broadband Metamaterials in Electromagnetics: Technology and
  Applications}.
\newblock Pan Stanford Publishing, 2016.

\bibitem{Liu}
Z.~Liu, X.~Zhang, Y.~Mao, Y.Y. Zhu, Z.~Yang, C.T. Chan, and P.~Sheng.
\newblock Locally resonant sonic materials.
\newblock {\em Science}, 289(5485):1734--1736, 2000.

\bibitem{craster2012}
R.~V. Craster and S.~Guenneau.
\newblock {\em Acoustic Metamaterials: Negative Refraction, Imaging, Lensing
  and Cloaking}.
\newblock London: Springer, 2012.

\bibitem{elastic_book}
R~Craster and S~Guenneau.
\newblock {\em World Scientific Handbook of Metamaterials and Plasmonics:
  {V}olume 2: Elastic, Acoustic and Seismic Metamaterials}.
\newblock World Scientific, 2017.

\bibitem{brule}
S.~{Br\^ul\'e}, E.~H. Javelaud, S.~Enoch, and S.~Guenneau.
\newblock Experiments on seismic metamaterials: Molding surface waves.
\newblock {\em Phys. Rev. Lett.}, 112:133901, 2014.

\bibitem{mouldin_waves}
T.~Antonakakis, R.~V. Craster, and S.~Guenneau.
\newblock Moulding and shielding flexural waves in elastic plates.
\newblock {\em Euro. Phys. Lett.}, 105:54004, 2014.

\bibitem{djafar}
D. Torrent, Y. Pennec, and B. Djafari-Rouhani.
\newblock Omnidirectional refractive devices for flexural waves based on graded
  phononic crystals.
\newblock {\em J. Appl. Phys.}, 116(22):224902, 2014.

\bibitem{achaoui_seismic}
Y.~Achaoui, T.~Antonakakis, S.~Brule, R.~Craster, S.~Enoch, and S.~Guenneau.
\newblock Clamped seismic metamaterials: Ultra-low frequency stop bands.
\newblock {\em New J. Phys.}, 19:063022, 2017.

\bibitem{miniaci_seismic}
M. Miniaci, A. Krushynska, F. Bosia, and N.~M Pugno.
\newblock Large scale mechanical metamaterials as seismic shields.
\newblock {\em New J. Phys.}, 18(8):083041, 2016.

\bibitem{DAlessandro}
L. D'Alessandro, R. Ardito, F. Braghin, A. Corigliano, \newblock Low frequency 3D ultra-wide vibration attenuation via elastic metamaterial.
\newblock {\em Sci. Rep.}, 9:8039, 2019. 

\bibitem{DAlessandro1}
L. D'Alessandro, B. Bahr, L. Daniel, D. Weinstein, R. Ardito \newblock Shape optimization of solid–air porous phononic crystal slabs with widest full 3D bandgap for in-plane acoustic waves.
\newblock {\em Journal of Computational Physics}, Vol. 344, 2017.

\bibitem{eleni2}
V.K. Dertimanis, I.A. Antoniadis, and E.N. Chatzi.
\newblock Feasibility analysis on the attenuation of strong ground motions
  using finite periodic lattices of mass-in-mass barriers.
\newblock {\em J. Engng Mech.}, 142(9):04016060, 2016.

\bibitem{finocchio}
G.~Finocchio, O.~Casablanca, G.~Ricciardi, U.~Alibrandi, F.~Garescì,
  M.~Chiappini, and B.~Azzerboni.
\newblock Seismic metamaterials based on isochronous mechanical oscillators.
\newblock {\em Appl. Phys. Lett.}, 104(19):191903, 2014.

\bibitem{carta_giorgio}
G.~Carta, I.S. Jones, N.~V. Movchan, A.~B. Movchan, and M.~J. Nieves.
\newblock Gyro-elastic beams for the vibration reduction of long flexural
  systems.
\newblock {\em Proc. R. Soc. Lond. A}, 473:2017.0136, 07 2017.


\bibitem{cocacola}
F.~Lemoult, M.~Fink, and G.~Lerosey.
\newblock Acoustic resonators for far-field control of sound on a subwavelength
  scale.
\newblock {\em Phys. Rev. Lett.}, 107:064301, 2011.

\bibitem{fano}
A.~E. Miroshnichenko, S.~Flach, and Y.~S. Kivshar.
\newblock Fano resonances in nanoscale structures.
\newblock {\em Rev. Mod. Phys.}, 82:2257--2298, 2010.

\bibitem{ColombiSensing}
A. Colombi, V. Ageeva, R. Smith, R. Patel, M. Clark,
  R. Craster, A. Clare, S. Guenneau, and P. Roux.
\newblock Enhanced sensing and conversion of ultrasonic rayleigh waves by
  elastic metasurfaces.
\newblock {\em Sci. Rep.}, 7:6750, 02 2017.

\bibitem{andrea6}
A.~Colombi.
\newblock Resonant metalenses for flexural waves.
\newblock {\em J.\ Acoust.\ Soc.\ Am.}, 140(5):EL423, 2016.

\bibitem{Matlack26072016}
{K.H. Matlack, A. Bauhofer, S. Kr{\"o}del, A. Palermo and C. Daraio,
  }
\newblock {Composite 3D-printed metastructures for low-frequency and broadband
  vibration absorption}.
\newblock {\em Proc. Natl. Acad. Sci.}, 113(30):8386--8390, 2016.

\bibitem{celli1}
D. Cardella, P. Celli, and S. Gonella.
\newblock Manipulating waves by distilling frequencies: a tunable shunt-enabled
  rainbow trap.
\newblock {\em Smart Materials and Structures}, 25(8):085017, 2016.

\bibitem{ErturkHarvesting}
A. Erturk and N. Elvin.
\newblock {\em Advances in Energy Harvesting Methods}.
\newblock Springer, 2013.

\bibitem{ErturkPiezoBook}
A. Erturk and D.~J. Inman.
\newblock {\em Piezoelectric Energy Harvesting}.
\newblock John Wiley and Sons, Ltd, 2011.

\bibitem{StevenAnton}
S.~R. Anton and H.~A. Sodano.
\newblock A review of power harvesting using piezoelectric materials
  (2003-2006).
\newblock {\em Smart Materials and Structures}, 16(3):R1--R21, May 2007.

\bibitem{RiazAhmed}
R. Ahmed, F. Mir, and S. Banerjee.
\newblock A review on energy harvesting approaches for renewable energies from
  ambient vibrations and acoustic waves using piezoelectricity.
\newblock {\em Smart Materials and Structures}, 26(8):085031, July 2017.

\bibitem{Carrara}
M. Carrara, M.~R. Cacan, J.~Toussaint, M. Leamy, M. Ruzzene, and
  A.~Erturk.
\newblock Metamaterial-inspired structures and concepts for elastoacoustic wave
  energy harvesting.
\newblock {\em Smart Materials and Structures}, 22:065004, 04 2013.

\bibitem{HangHarvesting}
H. Lv, X. Tian, M.~Y. Wang, and D. Li.
\newblock Vibration energy harvesting using a phononic crystal with point
  defect states.
\newblock {\em Appl. Phys. Lett.}, 102:034103, January 2013.

\bibitem{LiangWu}
L.-Y. Wu, L.-W. Chen, and C.-M. Liu.
\newblock Acoustic energy harvesting using resonant cavity of a sonic crystal.
\newblock {\em Appl. Phys. Lett.}, 95:013506, June 2009.

\bibitem{Gonnella}
S. Gonella, A.~C. To, and W.~K. Liu.
\newblock Interplay between phononic bandgaps and piezoelectric microstructures
  for energy harvesting.
\newblock {\em J. Mech. Phys. Solids}, 57:621633, November 2009.

\bibitem{Tol}
S.~Tol, F.~L. Degertekin, and A.~Erturk.
\newblock Phononic crystal luneburg lens for omnidirectional elastic wave
  focusing and energy harvesting.
\newblock {\em Appl. Phys. Lett.}, 111:013503, July 2017.

\bibitem{Tol2}
S.~Tol, F.~L. Degertekin, and A.~Erturk.
\newblock Gradient-index phononic crystal lens based enhancement of elastic
  wave energy harvesting.
\newblock {\em Appl. Phys. Lett.}, 109:064902, May 2016.

\bibitem{AhmadLens}
A. Zareei, A. Darabi, M.~J. Leamy, and M.-R. Alam.
\newblock Continuous profile flexural {GRIN} lens: {F}ocusing and harvesting
  flexural waves.
\newblock {\em Appl. Phys. Lett.}, 112:023901, January 2018.

\bibitem{Sugino}
C~Sugino and A~Erturk.
\newblock Analysis of multifunctional piezoelectric metastructures for
  low-frequency bandgap formation and energy harvesting.
\newblock {\em J. Phys. D: Appl. Phys}, 51:215103, May 2018.

\bibitem{CVETICANIN201789}
L.~Cveticanin and M.~Zukovic.
\newblock Negative effective mass in acoustic metamaterial with nonlinear
  mass-in-mass subsystems.
\newblock {\em Communications in Nonlinear Science and Numerical Simulation},
  51:89 -- 104, 2017.
  
\bibitem{carbonara}
A. Casalotti, S. El-Borgi, and W. Lacarbonara.
\newblock Metamaterial beam with embedded nonlinear vibration absorbers.
\newblock {\em International Journal of Non-Linear Mechanics}, 98:32--42, 2018.

\bibitem{colombi16a}
A.~Colombi, D.~Colquitt, P.~Roux, S.~Guenneau, and R.~V. Craster.
\newblock A seismic metamaterial: The resonant metawedge.
\newblock {\em Sci. Rep.}, 6:27717, 2016.

\bibitem{wu}
T-T. Wu, Z.-G. Huang, T.-C. Tsai, and T.-C. Wu.
\newblock Evidence of complete band gap and resonances in a plate with periodic
  stubbed surface.
\newblock {\em Appl. Phys. Lett.}, 93(11):111902, 2008.

\bibitem{pennecchia}
Y.~Pennec, B.~Djafari-Rouhani, H.~Larabi, J.~O. Vasseur, and A.~C.
  Hladky-Hennion.
\newblock Low-frequency gaps in a phononic crystal constituted of cylindrical
  dots deposited on a thin homogeneous plate.
\newblock {\em Phys. Rev. B}, 78:104105, Sep 2008.
  
\bibitem{younes2011}
Y.~Achaoui, A.~Khelif, S.~Benchabane, L.~Robert, and V.~Laude.
\newblock Experimental observation of locally-resonant and bragg band gaps for
  surface guided waves in a phononic crystal of pillars.
\newblock {\em Phys. Rev. B}, 83(10):10401, 2011.

\bibitem{matthieu}
M.~Rupin, F.~Lemoult, G.~Lerosey, and P.~Roux.
\newblock Experimental demonstration of ordered and disordered multi-resonant
  metamaterials for {L}amb waves.
\newblock {\em Phys. Rev. Lett.}, 112:234301, 2014.

\bibitem{perkins86a}
N.~C. Perkins and C.~D. Mote.
\newblock Comments on curve veering in eigenvalue problems.
\newblock {\em J. Sound Vib.}, 106:451--463, 1986.

\bibitem{landau58a}
L.~D. Landau and E.~M. Lifshitz.
\newblock {\em Quantum Mechanics non-relativistic theory}.
\newblock Pergamon Press, 1958.

\bibitem{andrea_tree}
A.~Colombi, P.~Roux, S.~Guenneau, P.~Gueguen, and R.cV. Craster.
\newblock Forests as a natural seismic metamaterial: Rayleigh wave bandgaps
  induced by local resonances.
\newblock {\em Sci. Rep.}, 5(5):19238, 2016.

\bibitem{kaina15a}
N.~Kaina, F.~Lemoult, M.~Fink, and G.~Lerosey.
\newblock Negative refractive index and acoustic superlens from multiple
  scattering in single negative metamaterials.
\newblock {\em Nature}, 525:77--81, 2015.

\bibitem{ColombiMechRew}
A. Colombi, R. Craster, D. Colquitt, Y. Achaoui, S. Guenneau, P. Roux, and M.  Rupin.
\newblock Elastic wave control beyond band-gaps: Shaping the flow of waves in
  plates and half-spaces with subwavelength resonant rods.
\newblock {\em Frontiers in Mechanical Engineering}, 3, 05 2017.

\bibitem{kittel}
C. Kittel.
\newblock {\em Introduction to Solid State Physics}.
\newblock Wiley, Hoboken, NJ, 2005.

\bibitem{comsol}
{COMSOL}.
\newblock {\em {www}.comsol.com}, 2012.

\bibitem{williams}
E.~G. Williams, P. Roux, M. Rupin, and W.~A. Kuperman.
\newblock Theory of multiresonant metamaterials for ${A}_{0}$ lamb waves.
\newblock {\em Phys. Rev. B}, 91:104307, 2015.

\bibitem{hess2007}
K.~L. Tsakmakidis, A.~D. Boardman, and O. Hess.
\newblock {Trapped rainbow} storage of light in metamaterials.
\newblock {\em Nature}, 450:397--401, 2007.

\bibitem{Gafforelli}
G. Gafforelli, R. Ardito, and A. Corigliano.
\newblock Improved one-dimensional model of piezoelectric laminates for
  energy harvesters including three dimensional effects.
\newblock {\em Composite Structures}, 127:369--381, 2015.

\bibitem{Gafforelli1}
R. Ardito, A. Corigliano, G. Gafforelli, C. Valzasina, F. Procopio, R. Zafalon.
\newblock Advanced model for fast assessment of piezoelectric micro energy harvesters.
\newblock {\em Frontiers in Materials}, Vol. 3, 2016.

\bibitem{AbsorbingBC}
P.~Rajagopal, M.~Drozdz, E.A Skelton, Lowe M.~J S, and R.V. Craster.
\newblock On the use of absorbing layers to simulate the propagation of elastic
  waves in unbounded isotropic media using commercially available finite
  element packages.
\newblock {\em {NDT} E International}, 51:30 -- 40, 2012.












































































































































































































































































\end{thebibliography}
\end{document}